# Electronic, magnetic and galvanomagnetic properties of Co-based Heusler alloys: possible states of a half-metallic ferromagnet and spin gapless semiconductor

A. A. Semiannikova[1*], Yu. A. Perevozchikova[1], V. Yu Irkhin[1], E. B. Marchenkova[1], P. S. Korenistov[1], and V. V. Marchenkov[1,2*]

[1] 620108, M.N. Mikheev Institute of Metal Physics, Ekaterinburg, Russia
[2] 620002, Ural Federal University, Ekaterinburg, Russia

[*] e-mail: semiannikova@imp.uran.ru, march@imp.uran.ru

**Abstract**
Parameters of the energy gap and, consequently, electronic, magnetic and galvanomagnetic properties in different $X_2YZ$ Heusler alloys can vary quite strongly. In particular, half-metallic ferromagnets (HMFs) and spin gapless semiconductors (SGSs) with almost 100% spin polarization of charge carriers are promising materials for spintronics. The changes in the electrical, magnetic and galvanomagnetic properties of the $Co_2Y$Si ($Y$ = Ti, V, Cr, Mn, Fe) and $Co_2$Mn$Z$ Heusler alloys ($Z$ = Al, Si, Ga, Ge) in possible HMF and/or SGS states were followed and their interconnection was established. Significant changes in the values of the magnetization and residual resistivity were found. At the same time, the correlations between the changes in these electronic and magnetic characteristics depending on the number of valence electrons and spin polarization are observed.

**Introduction**
The motivation is that half-metallic ferromagnets (HMFs) and spin gapless semiconductors (SGSs) are promising materials for spintronics with almost 100% spin polarization of charge carriers. They possess a gap near the Fermi energy for the current carriers with spin down. However, for the opposite spin projection, these materials have a difference: the energy gap absences in HFMs, while it being zero in SGSs Refs. 1, 2.

It is not easy to unambiguously determine the conditions for the occurrence of HMF-states and, especially, SGS in practice. However, the observation of HMF and SGS-states was proposed in materials based on the $Co_2Y$Si ($Y$ = Fe, Mn), $Mn_2$CoAl and CoFeMnSi Heusler alloys and close to 100% spin polarization in Refs. 3-7. The densities of electronic states near the Fermi level $E_F$ change strongly with variation of the $Y$- and $Z$-components in the $X_2YZ$ Heusler compounds, which appears in changes in the electronic, magnetic, and optical properties Refs. 8-15. Generally, $Y$ are 3d-transition metals, and $Z$ are elements of the III-V group of the Periodic Table in the stoichiometric formula of Heusler alloys. Therefore, the aim of the work is to follow the changes in the electrical, magnetic and galvanomagnetic properties and to establish their interconnection and basic behavior patterns of the $Co_2Y$Si ($Y$ = Ti, V, Cr, Mn, Fe) and $Co_2$Mn$Z$ Heusler alloys ($Z$ = Si, Al, Ga, Ge) in the HMF- and/or SGS-states.

**Experimental**
Polycrystalline $Co_2Y$Si ($Y$ = Ti, V, Cr, Mn, Fe) alloys were prepared in an induction furnace in a purified argon atmosphere. Then, the obtained $Co_2$VSi, $Co_2$CrSi, and $Co_2$FeSi ingots were annealed at 1100°C for 3 days and quenched, whereas the $Co_2$TiSi and $Co_2$MnSi alloys were annealed at 800°C for 9 days, followed by cooling to the room temperature. Polycrystalline $Co_2$Mn$Z$ ($Z$ = Al, Si, Ga, Ge) alloys were prepared by arc melting methods in a purified argon atmosphere and annealed at 800 K during 48 h.

An elemental analysis was carried out by using a FEI Company Quanta 200 scanning electron microscope equipped with an EDAX X-ray microanalysis unit at the Collaborative Access Center «Testing Center of Nanotechnology and Advanced Materials» of the IMP UB RAS. The elemental composition was determined in at least three regions selected at three different points of the sample. The deviation from a stoichiometric composition was revealed to be insignificant in all the samples.

The electrical resistivity was measured by using a standard four-probe method in the temperature range from 0 to 300 K according to Ref. 16. The field dependences of the magnetization $M(H)$ $T = 4.2$ K were measured at in magnetic fields up to 50 kOe. The magnetization was measured using an MPMS-XL-5 SQUID magnetometer. A method for preparing samples and measuring magnetization are described in Ref. 17.

**Results and discussion**

Fig. 1 shows the X-ray diffraction patterns of $Co_2Y$Si ($Y$ = Ti, V, Cr, Mn, Fe) and $Co_2MnZ$ ($Z$ = Al, Si, Ga, Ge) alloys at room temperature. All alloys are seen to have the structure of the Heusler phase $L2_1$, which is evidenced by the presence of Bragg superstructure reflections of the 111 and 200 types. The lattice parameters presented in Tables I and II were determined. The experimental results are comparable with the data given in the literature, e.g. Ref. 18, see Tables I and II.

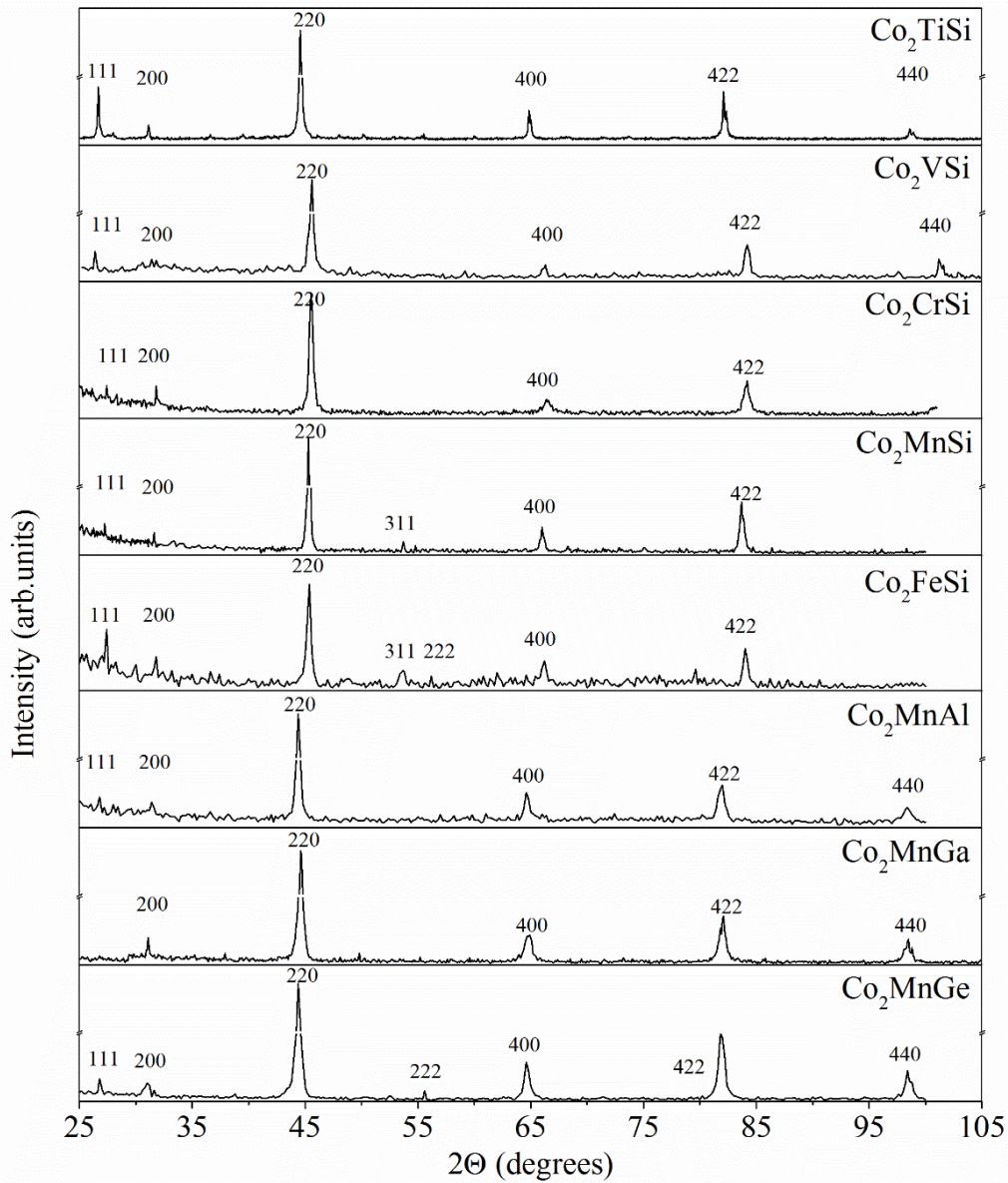

*Figure 1 - X-ray diffraction patterns of Co2MnZ (Z = Al, Ga, Ge, Sn) and Co2YSi (Y = Ti, V, Cr, Mn, Fe) alloys*

The electrical resistivity was measured in a wide temperature range from 4.2 to 300 K. Figs. 2 and 3 show that the $\rho(T)$ of the $Co_2VSi$, $Co_2CrSi$ and $Co_2MnAl$ are established to tend to saturation, while the $\rho(T)$ dependences for $Co_2TiSi$, $Co_2FeSi$, $Co_2MnGa$, $Co_2MnSi$ and $Co_2MnGe$ are linear at temperatures above 100 K.

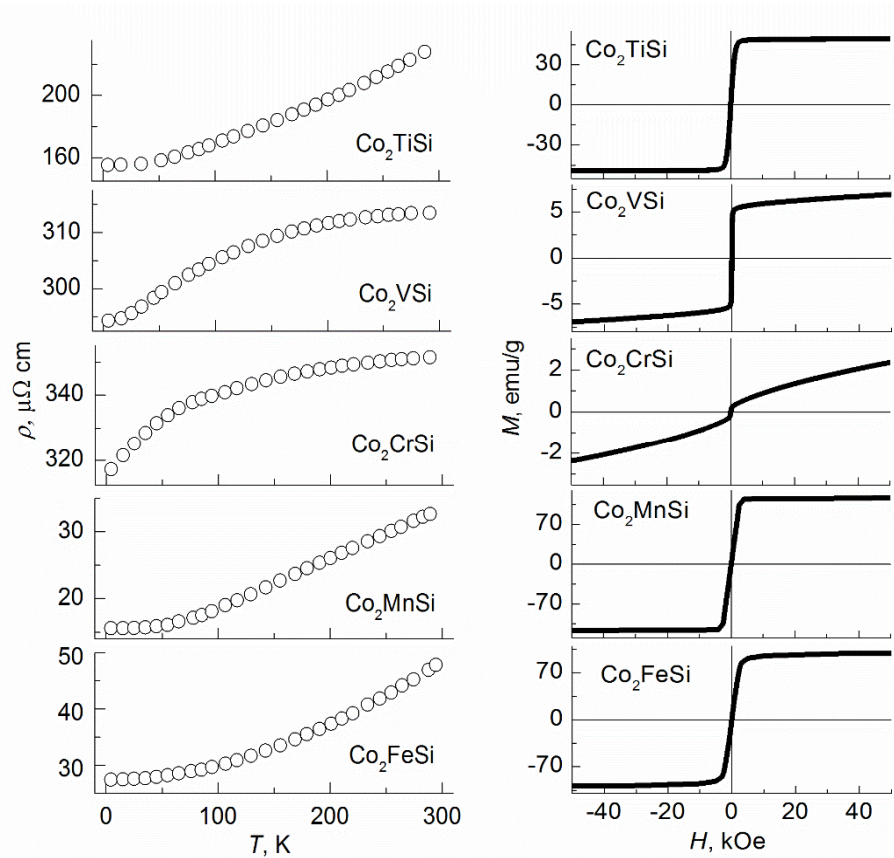

Figure 2— Temperature dependences of the electrical resistivity and field dependences of the magnetization at T = 4.2 K for the Co$_2$YSi system

The residual resistivity $\rho_0$ was determined at the temperature $T$ = 4.2 K for the Co$_2$YSi ($Y$ = Ti, V, Cr, Mn, Fe) system, Co$_2$MnAl and Co$_2$MnSi, the $\rho_0$ for the Co$_2$MnGa and Co$_2$MnGe was found by extrapolation the temperature dependences of resistivity to $T$ = 4.2 K.

Tables I and II contain values of the residual resistivity $\rho_0$, saturation magnetization $M_s$, lattice parameters, spin polarization of current carriers.

Table I- Number of valence electrons, the experimental lattice parameter, residual resistivity, saturation magnetization and spin polarization for the Co$_2$YSi system

| Compound | z | $r_0$, μΩcm | $M_S$, emu/g | P, % Refs. 3, 7, 19, 20 | $a_{exp}$, Å | a, Å Ref. 18 |
|---|---|---|---|---|---|---|
| Co$_2$TiSi | 26 | 155 | 48.8 | 24 | 5.747 | 5.743 |
| Co$_2$VSi | 27 | 294 | 5.9 | 35 | 5.650 | 5.657 |
| Co$_2$CrSi | 28 | 318 | 1.0 | 80 | 5.640 | 5.647 |
| Co$_2$MnSi | 29 | 16 | 114.2 | 93 | 5.660 | 5.645 |
| Co$_2$FeSi | 30 | 27 | 96.7 | 57 | 5.640 | 5.640 |

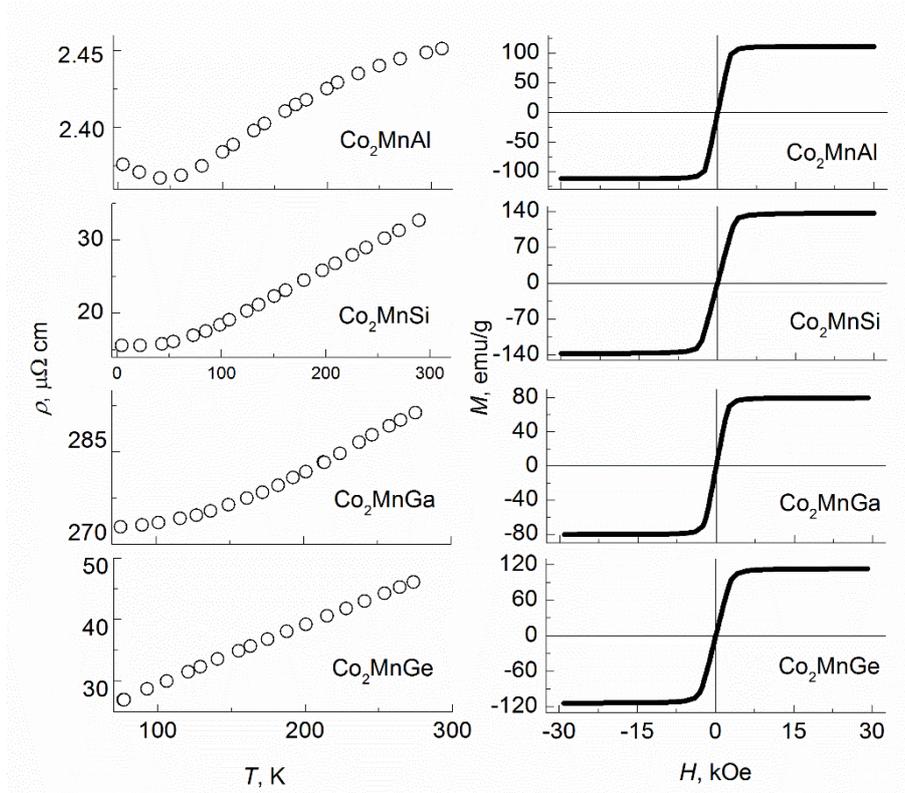

*Figure 3 — Temperature dependences of the electrical resistivity and field dependences of the magnetization at T = 5 K for the Co$_2$MnZ system*

Figs. 2 and 3 demonstrate that the magnetization curves $M(H)$ have a form characteristic of ordinary ferromagnets. These dependences do not demonstrate rapid saturation for Co$_2$VSi and Co$_2$CrSi which seem to be weak itinerant magnets rather than strong half-metallic ferromagnets.

The Co$_2$VSi, Co$_2$CrSi and Co$_2$MnGa alloys demonstrate relatively large residual resistivity $\rho_0$ varying from 272 to 318 μΩ×cm, simultaneously, the Co$_2$VSi and Co$_2$CrSi possess relatively low values of the saturation magnetization $M_s$. The occurrence of such peculiarities may indicate unique features near the Fermi level $E_F$, e.g., the proximity to the SGS-state with a small gap and low activation energy. On the contrary, Co$_2$MnSi, Co$_2$FeSi and Co$_2$MnAl have large $M_s$ and small $\rho_0$ values, therefore, the HMF-state seems to be observed.

Significant changes in the values of these properties are found. At the same time, the interconnection is observed between the changes in these electronic and magnetic characteristics depending on the number of valence electrons.

*Table II – The experimental lattice parameter, residual resistivity, saturation magnetization and spin polarization for the Co$_2$MnZ system*

| Compound | $r_0$, μΩcm | $M_S$, emu/g | $P$, % Ref. 21 | $a_{exp}$, Å | $a$, Å Ref. 18 |
|---|---|---|---|---|---|
| Co$_2$MnAl | 2.4 | 110.7 | 65.2 | 5.765 | 5.749 |
| Co$_2$MnSi | 16 | 132.2 | 100 | 5.660 | 5.645 |
| Co$_2$MnGa | 272 | 79.9 | 63.4 | 5.760 | 5.767 |
| Co$_2$MnGe | 24 | 108.7 | 100 | 5.760 | 5.749 |

Fig. 4 demonstrates temperature dependences of the magnetization at $H = 30$ kOe for the Co$_2Y$Si and Co$_2$Mn$Z$ systems.

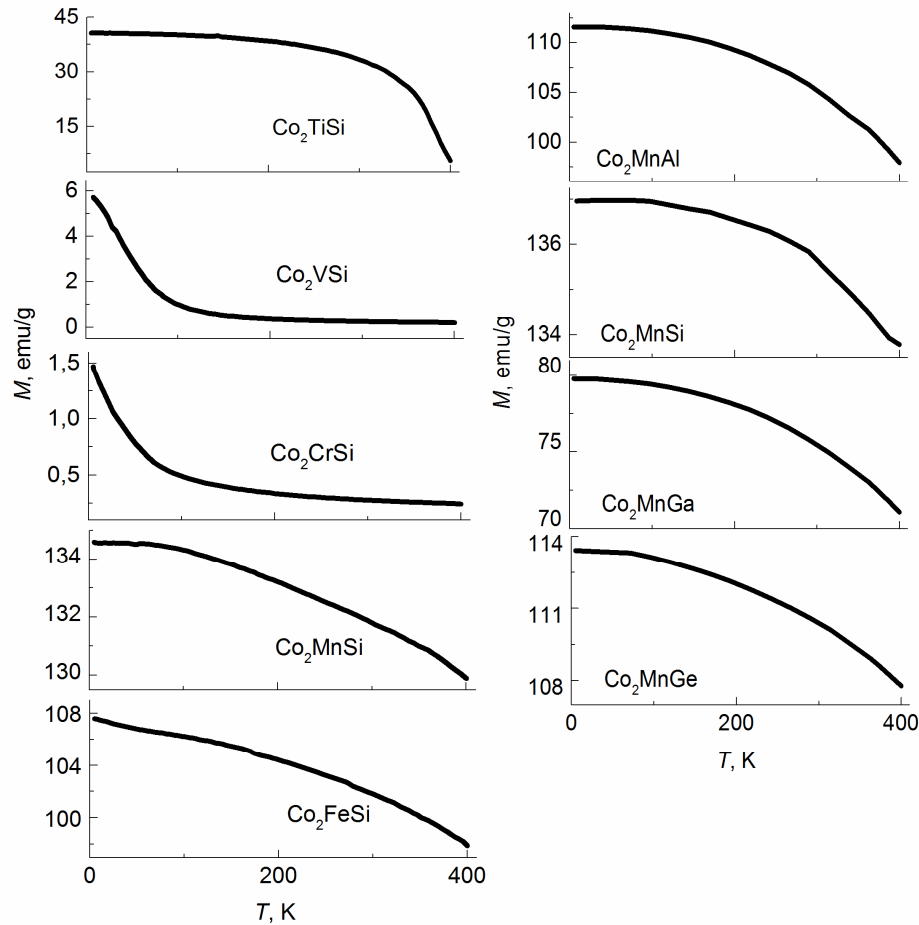

Figure 4 - Temperature dependences of the magnetization at H = 30 kOe for the Co$_2Y$Si (on the left) and Co$_2$MnZ (on the right) systems

Fig. 5 shows the calculated and experimental values of the coefficients of spin polarization of current carriers for the Co$_2Y$Si system taken from Refs. 3, 7, 19, 20 on the number of valence electrons $z$. It seems very interesting to analyze the behavior of the spin polarization with a change in the number of valence electrons and compare it with the data of the residual resistivity and the saturation magnetization. Fig. 5 shows a good correlation between the experimental data obtained in this work and the coefficient of current carriers spin polarization $P$ from literature data. When the number of valence electrons changes, the spin polarization peaks at z = 28 and z = 29 for Co$_2$CrSi and Co$_2$MnSi.

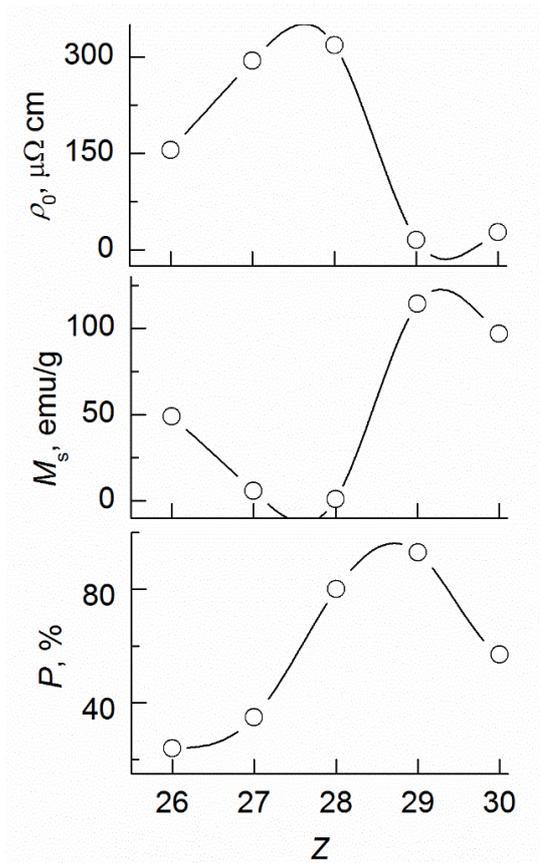

*Figure 5 – The correlation with spin polarization for the $Co_2YSi$ system*

Fig. 6 presents the correlation between obtained experimental data and the coefficients of current carriers spin polarization for the $Co_2MnZ$ system Ref. 21, when Z-component varies from Al to Ge according to their atomic mass. Thus Z-component variations change considerably atomic potentials and physical properties. Methods for determining spin polarization: VASP + GGA functional Ref. 19, 20, Point Contact Andreev Reflection (PCAR) spectroscopy Ref. 7, Spin resolved ultraviolet-photoemission spectroscopy (SRUPS) Ref. 3, GGA functional Ref. 21.

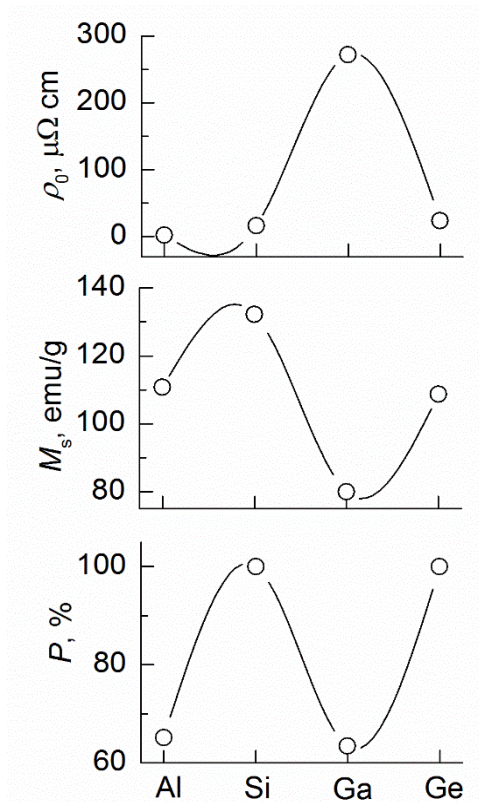

*Figure 6 – The correlation with spin polarization for the Co$_2$MnZ system*

A comparison of Figs. 5 and 6 indicates that the correlation is observed between the properties both in the Co$_2Y$Si and Co$_2$MnZ systems of Heusler alloys. The spin polarization values are consistent with the previously stated assumption that the Co$_2$MnSi alloy has a high spin polarization, which can be used for choosing promising materials for spintronic devices.

**Conclusions**
A number of features of electronic and magnetic properties can indicate peculiarities of the electron energy spectrum, i.e. HMF- and/or SGS-states. In some cases, the available experimental techniques do not allow to unambiguously distinguish between these situations.

New information has been obtained on the features of the electronic and magnetic characteristics of the Co$_2Y$Si ($Y$ = Ti, V, Cr, Mn, Fe) and Co$_2$MnZ ($Z$ = Al, Si, Ga, Ge) Heusler alloys, which may be useful in choosing optimal materials for spintronic devices.

Investigations of magnetization enable one to prove half-metallic nature of ferromagnetism and verify results of band structure calculations.

**Acknowledgments**
The work was performed within the framework of the state assignment of the Ministry of Science and Higher Education of Russia (the themes "Spin", No. AAAA-A18-118020290104-2-2 and "Quantum" No. AAAA-A18-118020190095-4) with partial support from the RFBR (projects No. 18-02-00739 and 20-32-90065) and the Government of the Russian Federation (Decree No. 211, Contract No. 02.A03.21.0006).

**Data availability**
The data that support the findings of this study are available from the corresponding author upon reasonable request.